# Ultra-low Power Domain Wall Device for Spin-based Neuromorphic Computing

Durgesh Kumar[1], Chung Hong Jing[2], Chan JianPeng[1], Tianli Jin[1], Lim Sze Ter[2], Rachid Sbiaa[3], and S.N. Piramanayagam[1#]

[1]School of Physical and Mathematical Sciences, Nanyang Technological University, 21 Nanyang Link, Singapore, 637371

[2]Institute of Materials Research and Engineering, A*STAR, 2 Fusionopolis Way, Innovis, Singapore, 138634

[3]Department of Physics, Sultan Qaboos University, Muscat, Oman

[#]Email: prem@ntu.edu.sg

**Abstract**

Neuromorphic computing (NC) is gaining wide acceptance as a potential technology to achieve low-power intelligent devices. To realize NC, researchers investigate various types of synthetic neurons and synaptic devices such as memristors and spintronic domain wall (DW) devices. In comparison, DW-based neurons and synapses have potentially higher endurance. However, for realizing low-power devices, DW motion at low energies — typically below pJ/bit—are needed. Here, we demonstrate domain wall motion at current densities as low as $10^6$ A/m$^2$ by tailoring the $\beta$-W spin Hall material. With our design, we achieve ultra-low pinning fields and current density reduction by a factor of $10^4$. The energy required to move the domain wall by a distance of about 20 μm is 0.4 fJ, which translates into energy consumption of 0.4 aJ/bit for a bit-length of 20 nm. With a meander domain wall device configuration, we have established a controlled DW motion for synapse applications and have shown the direction to make ultra-low energy spin-based neuromorphic elements.

Artificial intelligence (AI) is now widely being used, ranging from smartphone applications to self-driving cars. To overcome the high power consumption of AI applications, researchers are investigating neuromorphic computing (NC). Since the human brain operates at 20 W and performs intelligent tasks that supercomputers perform by consuming kilo-watts of power, NC (a term for brain-inspired computing) is expected to be energy efficient. For practical implementation of NC, researchers need to invent the electronic analogues of neurons (processing units of the brain) and synapses (memory units of the brain). Ferroelectric, resistive and magnetic materials have been proposed for designing synapse [1-3]. Researchers have also investigated spin-based neurons for character and voice recognition [4,5]. Despite all these researches, NC is at a primitive stage and one of the major problems facing the implementation of NC is the high power consumption of the synthetic synapses and neurons.

Amongst several devices for neuromorphic computing based on charge and spin, domain wall (DW) devices are one of the least power consuming [6,7]. However, most of the investigations on DW devices for NC are only based on simulation, and the experimental realization of the proposed device for NC is only developing [8-13]. For example, Jin et al. fabricated artificial pinning sites using an ion implantation method for achieving multi-resistance states with a purpose of using it as neuromorphic synapse [14]. In another study, multi-resistance states were obtained by increasing the width of the nanowire at certain spacing [15]. Borders et al. fabricated a Hall device and showed multiple Hall resistance states [16]. Recently, Sato et al. fabricated a *z*-shaped nanowire to achieve the DW oscillations at GHz frequencies for neuromorphic application [17].

For experimental realization of low-power neuromorphic computing based on DW devices, it is very important to shortlist a suitable driving force for the DWs. External magnetic field, which has been investigated as a stimulus for driving the DWs along a nanotrack [18], is relatively more power consuming and inapt for large-scale fabrication of DW devices on a single chip.

Spin-transfer torque, which can be decomposed into adiabatic and non-adiabatic torque, can move the DWs in ferromagnetic wire [19]. In the sole presence of adiabatic torque, the DW moves only above a threshold current density. The typical current densities are of the order of $10^{11}$ A/m$^2$, which translates to an energy of the order of ~1 µJ/bit for translation of DW by a distance of 20 nm [20].

The critical current density for driving the DWs in ferromagnetic (FM) wire can further be reduced using a spin-orbit torque (SOT) from the adjacent heavy metal (HM) layer of high spin-orbit coupling [3,21-26]. The efficiency of Slonczewski-like torque from the spin Hall effect depends on the conversion efficiency of the charge current to spin current [27]. Therefore, a larger spin-Hall angle (SHA or $\theta_{SH}$) material is a choice to enhance SOT efficiency and hence, to lower the current density for moving the DWs. Several HM materials such as Au, Nb, Mo, Hf, Pd, Pt, Ta, W, $W_{1-x}Ta_x$, PtMn have been investigated for enhancing $\theta_{SH}$ [28-31]. Among these materials thin W with metastable $\beta$-phase (A15 crystal structure) has been the choice of many research studies due to a relatively higher $\theta_{SH}$. An increment in the resistivity of $\beta$-W in a range of 100-200 µΩcm has been considered as the basis for increasing the SHA. Researchers have studied several methods viz., tuning of film deposition parameters, oxygen doping in W layer and alloying W with other heavy metals for increasing the SHA. Most of the studies reported the $\theta_{SH}$ value in a range of -0.2 to -0.6 [28-30,32-35]. However, resistivity at the higher end of 200 µΩcm is not desirable as this will lead to more power consumption [36]. Moreover, it may inherently add several other issues such as high Joule heating, higher intrinsic defects in FM layer and lower tunnel magnetoresistance (TMR) signal [37,38]. The typical current density for moving DW in the case of SOT is of the order of ~$10^{10}$ A/m$^2$, which translates to an energy of about ~1 nJ/bit for moving DW by a distance of 20 nm [39].

In this article, we have carried out material engineering and device designs to conceive low-power synapses and neurons and showed energy consumption in the order of aJ/bit. Our spin Hall material enables us to move DW at a current density, as low as $10^6$ A/m$^2$, which is 10000 times less than the values reported in the literature. We also report device designs that enable us to stop the DWs precisely, a desired feature for achieving non-stochastic synaptic devices.

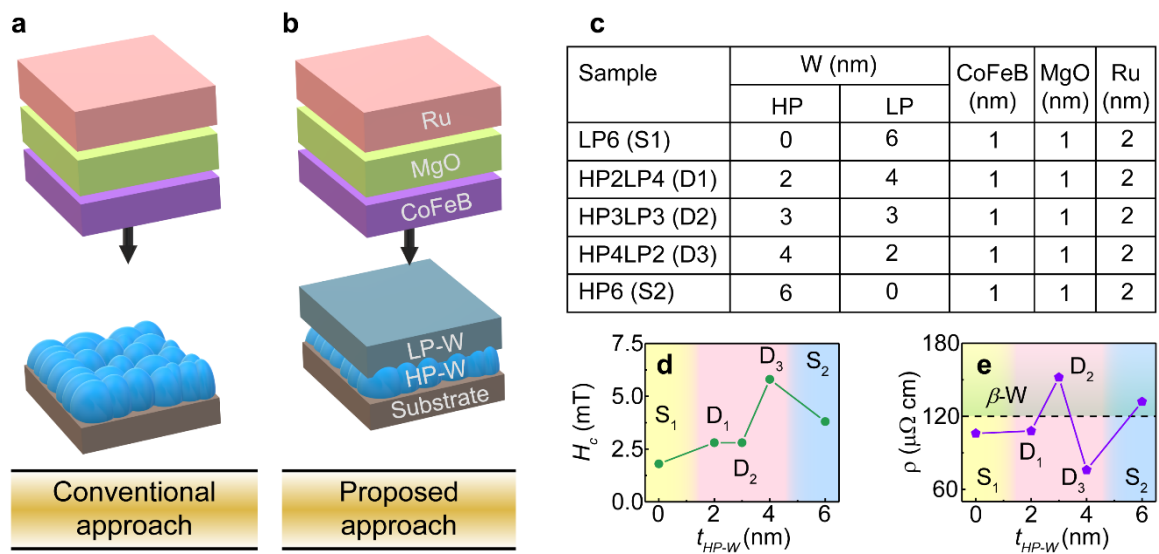

*Fig. 1. Illustration of stack structure and measured properties of thin film samples* (a) The schematics of the conventional approach, where high pressure (HP) β-W is used for achieving high spin Hall angle. The HP and low-power deposition of W results in a granular structure of W layer. The granularity may be transferred to the ferromagnetic (FM) layer and introduce defects, which cause the stochastic DW motion and higher critical current density [40]. (b) In the proposed idea, the FM layer and HP−W layer are separated by a low- pressure (LP) and low-power deposited W film. Insertion of this LP-W is expected to reduce the resistance of W stack and the defects in FM layer. (c) Table of sample codes and the thickness of each layer (d) The coercivity ($H_c$) of the films samples measured in the out-of-plane direction. All the samples showed a PMA and $H_c$ is the lowest for LP-W film, while the presence of HP-W increases the $H_c$. (e) The plot of stack resistivity for different films. A stack resistivity more than 120 μΩcm is generally considered as the baseline for the β phase. Most of the samples with HP-W show β phase, as indicated in the graph.

**Thin film investigations**

We made film stacks of the type shown in figure 1 (a, b). Sample shown in figure 1(a) illustrates the conventional approach followed by various groups [28-30,32-35], wherein a high SHA is achieved for very high-resistivity values. In contrast, we propose a unique approach shown in figure 1(b), where we utilize a dual W layer stack to achieve complementing properties. While the high-pressure W (HP-W) is useful for achieving high SHA, a low-pressure W (LP-W) thin film, is expected to keep the resistivity at low levels yet achieve more efficient DW motion. This material design was expected to solve the problems of power consumption, Joule heating and TMR. The details of the deposition conditions and sample characterization techniques are presented in methods section. The roughness data and hysteresis loops of the film stack are shown in the supplementary section.

All the samples exhibited square hysteresis loops in the out-of-plane (OOP) direction indicating the presence of a perpendicular magnetic anisotropy (PMA) at as-deposited state, which is generally attributed to the presence of W in $\beta$-phase [41]. The coercivity of the samples showed an increase in the presence of a layer with HP-W (figure 1(d)). As the presence of PMA is not a clear evidence to confirm the $\beta$-phase in W spin Hall layer, we performed the resistivity measurements on our samples (figure 1(e)). It has been reported that if the stack resistivity (W ($t_W$ nm)/CoFeB (1 nm)/ MgO/ Cap layer) is over 120 $\mu\Omega$cm, the W is expected to be $\beta$-phase [41]. For W in $\alpha$-phase, the resistivity of the stack is typically below 30 $\mu\Omega$cm. In samples with thin W layers, where XRD peaks of $\beta$-W could not be observed, this criterion is useful.

The resistivity of samples LP6 and HP6 was found to be 106 and 132 $\mu\Omega$cm respectively, which suggests that sample HP6 possesses the $\beta$-phase. The resistivity of the stacks with combined HP-W and LP-W layers (called dual W layer henceforth), showed an increase in the resistivity with the thickness of the HP-W layer and shows a maximum of 166 $\mu\Omega$cm for HP-W (2.5 nm)/LP-W (3.5 nm) (data point is presented in supplementary). The samples with HP-W (6 nm) and HP-W (3 nm)/LP-W (3 nm) (sample HP3LP3) exhibit resistivity more than the baseline value and hence possess the *β*-phase. Samples with stack resistivity very close to 120 μΩcm, LP6 and HP2LP4 as shown in figure 1(e), are expected to be in *α+β* mixture phase. For current-driven DW motion, we require dual W layer samples with optimized resistivity and spin Hall efficiency. Therefore, we have chosen samples HP3LP3, LP6 and HP6 for current-driven DW motion experiments. Samples LP6 and HP6 are investigated for reference.

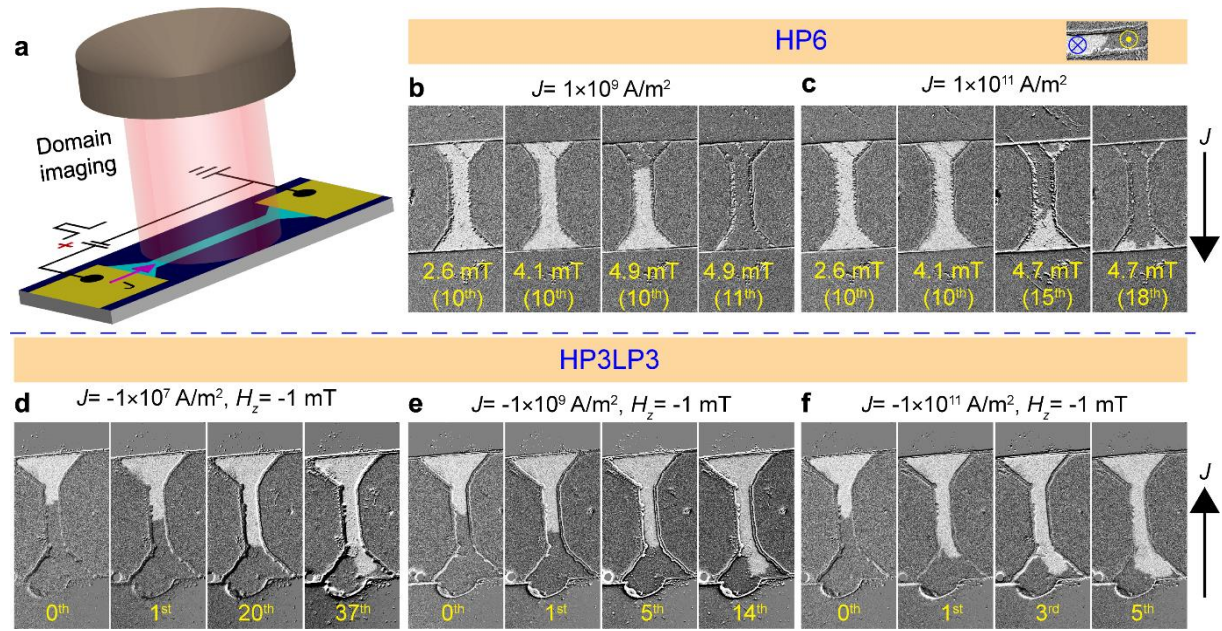

*Fig. 2*. ***Preliminary domain wall motion measurements*** *(a) Schematic of the domain wall (DW) motion measurement setup. (b-c) DW motion in device with conventional β-W spin-Hall layer (HP6) and (d-f) device with HP-W (3 nm)/LP-W (3 nm) spin Hall layer (HP3LP3). The images in (b-c) show that the domain wall motion in HP6 is rapid, and it does not occur until a large external magnetic field (closer to the coercivity) is applied. The images in (d-f) indicate that the domain wall motion occurs even for low current densities and the domain wall position depends on the number of pulses (pulse number is shown beneath the respective images) applied and the strength of external magnetic field.*

**Current-driven domain wall motion**

For current-driven DW motion experiments, we fabricated micro-wires of dimensions 50 μm × 10 μm on the above samples [20]. First, we compare the results obtained in sample HP6 (which has the conventional spin Hall layer) and the sample HP3LP3, which is the unique feature of this study.

The domain wall motion study involves (*i*) saturating the magnetization in the sample with a positive OOP magnetic field (~ 50 mT), (*ii*) applying a negative reversal field to nucleate a reverse domain, and (*iii*) applying current pulses to move the domain walls (in some cases, under an additional external magnetic field). In the case of HP6, the DW could not be nucleated and magnetization (along +*z* or -*z* axis) covering the whole microwire was considered as initial state. When the reversal field was increased gradually, a sharp switching of the magnetization covering the whole devices was obtained. Under the influence of a current pulse, the magnetic field required to observe the sharp magnetization switching of magnetization was reduced. However, even for very large current densities of $5 \times 10^{11}$ A/m$^2$, the switching was observed at magnetic fields of ~4.7 mT which is very close to the coercivity of the thin-film samples (please refer to figure 2 (b,c)).

In the case of devices of HP3LP3 sample, it was possible to nucleate DW by applying a much smaller negative field (-1.65 mT). Once a DW was nucleated, we applied current pulses along the microwire as shown in figure 2(d-f) and studied the motion of domain walls. Interestingly, the DW was found to move even for exceptionally small value of current density ~ $1 \times 10^7$ A/m$^2$. The domain walls moved by shorter distances and more pulses were required for travelling the total distance of 50 μm, at a low current density. Moreover, a small OOP magnetic field was applied during the measurements. The number of pulses required to displace domain walls (by ~50 μm) reduced significantly, when OOP magnetic field was increased by 0.1 mT. For the first time, DW motion at such a low current density has been observed, which is significantly new with a high potential for application. Consequently, we carefully studied the role of external magnetic field on the DW motion.

Figures 3 (a-d) show the snapshots of DW positions in HP3LP3 devices, at an ultra-low current density of $10^6$ A/m$^2$. It can be noticed from this figure that the domain wall starts moving even for a small external magnetic field of -1 mT. It can also be noticed that there is a correlation between the applied field direction (+/-z axis) and the current direction for achieving domain wall motion. To study this effect quantitatively, we carried out detailed measurements of domain wall velocity for different applied fields, current densities and their directions.

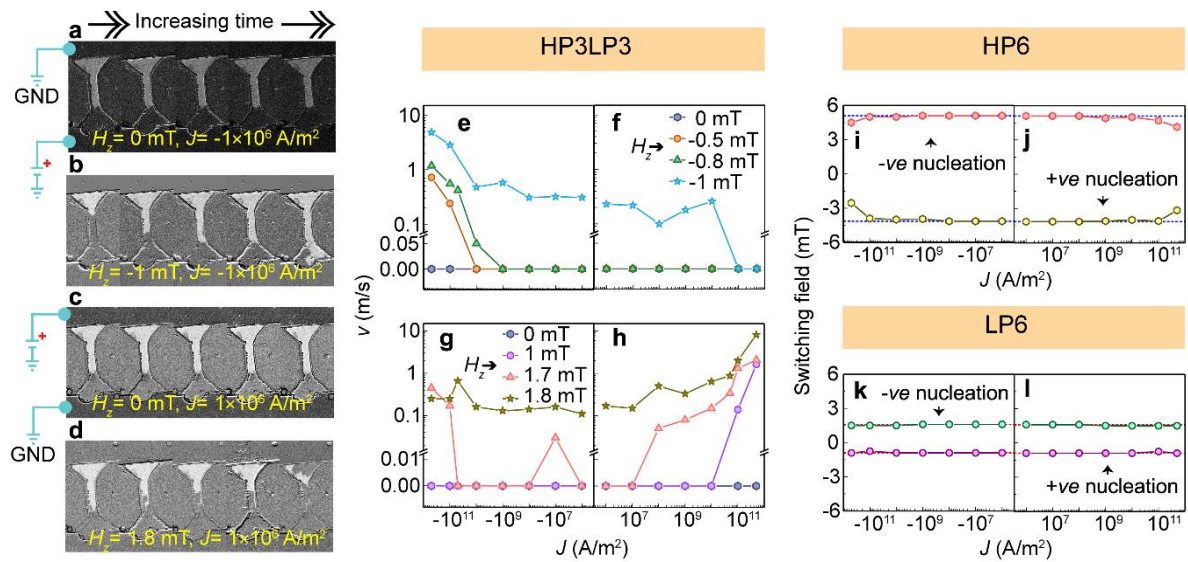

*Fig. 3. Current-induced domain wall motion investigations at different current densities and external magnetic fields* (a-d) Snapshots of DW positions in device with HP-W (3 nm)/LP-W (3 nm) spin Hall layers, for different values of external OOP magnetic fields and current densities. (e-h) DW velocity in device with HP-W (3 nm)/LP-W (3 nm) spin Hall layers, for different values of external magnetic fields and current densities. The magnetic field supports the current density in one direction and opposes in the reverse direction. The value of external magnetic fields at which the current density causes domain wall motion in device with (i-j) HP-W (6 nm) and (k-l) LP-W (6nm) spin Hall layers. DW velocity is difficult to measure in the single W layer samples, as the motion occurs only for field values nearer to the coercivity and the motion is swift (irrespective of the current density). The dotted lines represent the field at which switching happens in the absence of current. It can be noticed that the current density is of use only at $5 \times 10^{11}$ A/m$^2$ in sample HP6.

Figure 3 (e-h) show the domain wall velocity (*v*) as a function of current density (for two different directions, +*y* and -*y*) for various values of applied magnetic fields (in two different directions, +*z* and –*z*). Interestingly, the DW velocity increases with the current density only for a certain combination of current and magnetic field directions. For the negative current pulses under a negative magnetic field, DW moves even for a low current density of $10^6$ A/m$^2$ and a magnetic field of -1 mT. On the contrary, for positive current pulses and negative OOP magnetic field (figure 3(f)), the DW does not move at all until an OOP field -1 mT is applied. More importantly, the velocity decreases with the increase in current density in this case. Similarly, for positive current pulses and a positive magnetic field, the DW velocity increases with the current density, for magnetic fields above 1 mT. These results suggest that the negative OOP field assists the negative current and opposes the positive current and so on.

In this study, the direction of DW motion due to SOT is consistent with the previous reports [42,43]. When an ample amount of magnetization is orthogonal to the direction of spin current, the SOT acts like an effective field [44],

$$H_{SOT} = -\frac{\hbar \theta_{SH} J}{2|e|M_s t}(\hat{m} \times \hat{\sigma})$$

where $\hbar, \theta_{SH}, J, e, M_s$ and $t$ refer to the reduced the Planck constant, SHA, current density in HM layer, electronic charge, saturation magnetization and thickness of FM layer, respectively. Moreover, $\hat{m}$ and $\hat{\sigma}$ represent the direction of magnetization inside the DW and spin current. Therefore, when a current is flowing through the HM layer, it causes an OOP magnetic field and the direction of this OOP magnetic field depends on the direction of the current. In addition, the externally applied OOP magnetic field either assists or opposes the DW motion depending upon its relative direction with respect to OOP magnetic field due to SOT.

In our experiments, we have not applied any in-plane (IP) magnetic field and therefore, the DW motion is completely field-free (except for a negligible $H_z$). It may be argued that the domain wall motion at $10^6$ A/m$^2$ is achieved due to the presence of an external magnetic field of -1 mT and it may not be practical. Firstly, such a small $H_z$ of few mT can be easily implemented in devices using the stray field from the reference layers in an MTJ design. Moreover, it can also be seen from figure 3(i-j) that the DW motion is not observed in the other samples (LP6 or HP6) for such a small magnitude of external field, or smaller current densities. Therefore, this result is unique to the dual W layer design and is useful for designing low-energy synapse, where small current pulses can move the domain walls by short distances.

**Synaptic device and neuron investigations**

Being able to drive domain walls at low current densities is a significant result for synapse and neuron applications in low-energy spin-based neuromorphic computing. While there are many advances in terms of high-speed current-driven domain wall motion [23,24,26,43,45-48], DW motion at low-current densities is a top priority for NC. In figure 4(a,b), we illustrate how neurons and synapses may be designed using the domain-wall-based MTJs [49]. With neurons (figure 4(a)), the tunnel barrier is at the right edge of the DW device and the spike occurs when the domain wall is moved to the right extreme. The input spikes for this neuron comes from the other neurons, as in the brain. In the case of a synapse (figure 4(b)), the position of the domain wall MTJ changes based on the inputs received by the synapse. As a result, its resistance (weight) also changes. For brain, the driving force is electrochemical reactions. Whereas, the driving force here is SOT. For synapse, it is good to be able to design stochastic as well as non-stochastic devices. This could be achieved using our domain wall devices based on dual-layered sample HP3LP3.

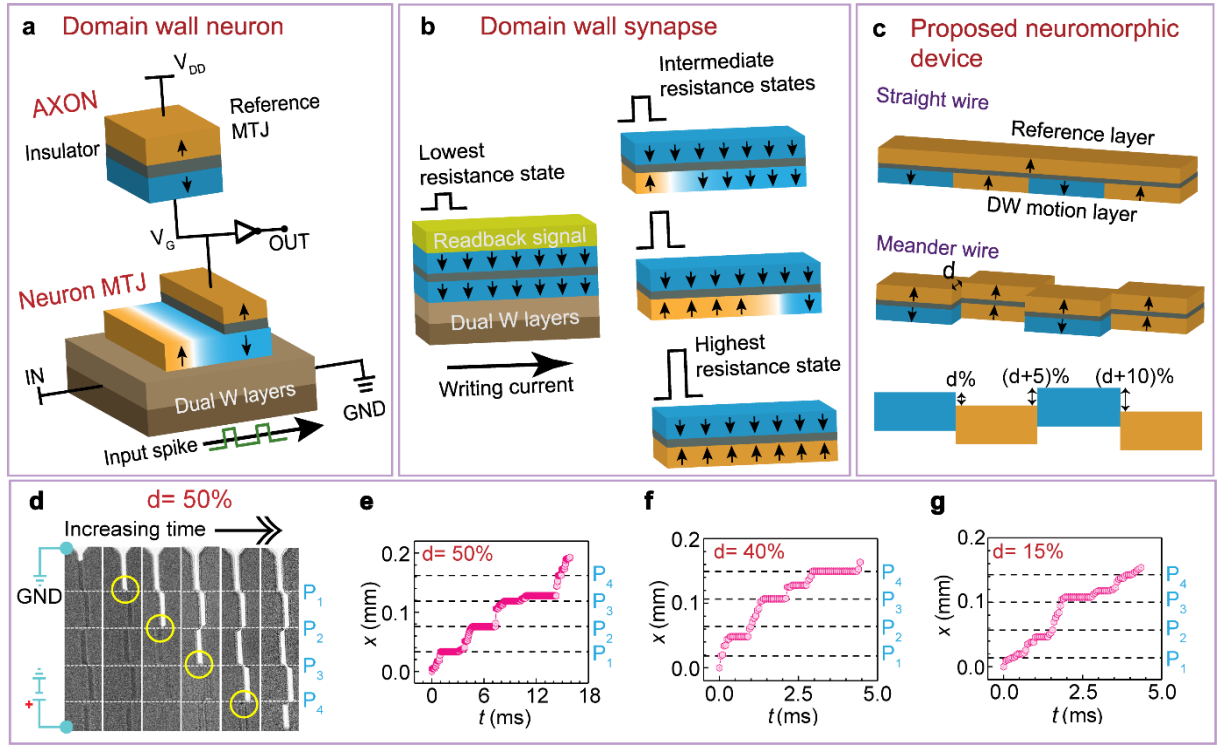

*Fig. 4. Preliminary investigations of synaptic properties of the domain wall devices* (a-b) Schematic illustration of the working principle of domain wall MTJ based neuron and synapses. The domain walls are moved by the input spikes from other neurons. The inputs are in the form of current and they induce SOT based domain wall motion. In the case of (a) neuron, the MTJ is at the right edge and the spike occurs when the domain wall is moved to the right edge. In the case of (b) synapse, the position of the domain wall MTJ changes based on the inputs the synapse receives from the other neurons. As a result, its resistance (weight) also changes. (c) The schematic of the meander wire for realizing the multiple resistance states. The reference nanowire (straight nanowire) is drawn for the comparison of the design. (d-g) The domain wall motion in devices without significant pinning "d= 15% and 40%" (useful for stochastic synapse) and with optimum pinning "d= 50%" (useful for non-stochastic synapse).

To achieve stochastic and non-stochastic synapses, we investigated straight microwires and also a new type of meander microwires. In the meander microwires (shown in figure 4(c)), the neighboring segments are offset by a distance *d*, expressed as a percentage of the width of the wire. The offset *d* determines the pinning strength. In our design, *d* also increases by 5% (5% of width of wire) in going from one pinning site to the next, in a particular direction.

To systematically study the effect of *d*, we made samples with various *d* such as 50%, 40% and 15%. Figure 4(d) shows the successful pinning of DW at all the pinning sites for *d* = 50%. The DW moves at current density values between $1\times10^6$ and $1\times10^7$ A/m² in the straight portions of meander wire. However, a higher *J* is needed to depin the DW from the pinning site. The magnitude of depinning current depends on the strength of the pinning site. For *d* of 40%, the DW pins at two pinning sites (offset= 50% and 55%) and does not pin at rest of the pinning sites (offset= 40% and 45%). The device with *d* = 15% does not exhibit pinning of the DW at the corners.

A potential design for synapse that has an MTJ with a TMR of 200% could use 8 levels. Such an MTJ could be 160 nm long, which means each bit occupying a length of 20 nm. From this study, the average energy consumed to cause domain wall movement by a distance of 20 nm is 0.6 aJ/bit. In comparison, the corresponding energy calculated from the other researchers are estimated at 400 fJ/bit [for W-Hf spin Hall layers] and nJ/bit [Ta spin Hall layer] [39]. The low energy consumption in our devices is arising from the dual-W spin Hall layer, which helps to reduce the intrinsic pinning field in addition to providing a reasonable spin Hall efficiency. The results demonstrated in this paper provide a path to achieve ultra-low energy neuromorphic spintronics.

**Conclusions**

The insertion of LP-W spin Hall layer between HP-W spin Hall layer and ferromagnetic layer has resulted in the reduction in the intrinsic pinning in the FM layer, without compromising the spin Hall angle. This, in turn, helped in reducing the current density for driving DWs by an order of $10^4$ than the values reported in the literature. The corresponding energy consumption can be reduced to 0.4 aJ/bit, which is significantly lower than all the reported studies. Moreover, we have proposed and demonstrated the design of meander nanowire for realizing the stochastic and non-stochastic DW synapses. The pinning probability can be tuned based on the geometrical design of these pinning centres. These observations provide the avenue for the ultra-energy efficient DW based neuromorphic computing devices.

**Methods:**

We have used DC/RF Singulus Timaris sputtering tool to deposit the samples, which had the following layer structure: "Si wafer/ W (6 nm)/ Co$_{40}$Fe$_{40}$B$_{20}$ (1 nm)/ MgO (~1 nm)/ Ru (2 nm)". The sample structures are shown in Table I. Several researchers have reported that a low deposition power ($P_W$) and high Ar gas pressure ($p_{Ar}$) (above 1 mTorr) are required during deposition of W thin films to achieve the $\beta$-phase in W, a prerequisite for achieving high SOT efficiency [32,50]. High $p_{Ar}$ of 3.63 mTorr is used to prepare high pressure (HP) W layer and a low $p_{Ar}$ of 0.78 mTorr for low pressure (LP) W layer at identical $P_W$ of 100 W. The absolute value of our sputtering power (100 W) appears to be larger than in the literature. However, the power density should be comparable, as the target size in Singulus Timaris sputtering system is large. Such a low power deposition resulted in a very low deposition rate of 0.03 nm/s. As shown in Table I, LP6 had only LP 6 nm W layer and HP6 had only HP 6 nm W layer. HP$_x$LP$_y$ layers had HP W layer at the bottom and LP W layer on top of the HP W layer. In all the stacks, the total thickness of the HM layer was fixed at 6 nm. The thickness values of other layers were kept the same in all the samples.

Table I. Stack structure and the thickness of the layers in different samples.

| Sample | Tungsten layer thickness (nm) | | CoFeB (nm) | MgO (nm) | Ru (nm) |
|---|---|---|---|---|---|
| | High pressure (3.63 mTorr) | Low pressure (0.78 mTorr) | | | |
| LP6 (S1) | 0 | 6 | 1 | 1 | 2 |
| HP2LP4 (D$_1$) | 2 | 4 | 1 | 1 | 2 |
| HP3LP3 (D2) | 3 | 3 | 1 | 1 | 2 |
| HP4LP2 (D$_3$) | 4 | 2 | 1 | 1 | 2 |
| HP6 (S2) | 6 | 0 | 1 | 1 | 2 |

We used atomic force microscopy (AFM) for estimating the surface roughness of all the films from area scan of 1 μm × 1 μm. We employed standard four-probe method for measuring the sheet resistance of the films. For this, we fabricated four pads using optical lithography and AJA sputtering tool of dimensions 300 μm × 300 μm at a centre to centre spacing of 0.25 cm. Subsequently, we applied a DC across the outermost electrodes and measured the voltage between inner electrodes. After a linear fitting of V *vs* I graph, we estimated the sheet resistance values. We have used a vibrating sample magnetometer (VSM) and alternating gradient magnetometer (AGM) for measuring in-plane (IP) and out-of-plane (OOP) magnetic hysteresis (M-H) loops and first-order reversal curves (FORC), respectively. Moreover, we have used magneto-optic Kerr microscopy for observing the magnetization reversal process of all the samples. We fabricated the DW and synaptic devices using optical lithography and ion milling process. We used a Keithley 6221 current source for applying current pulses and a Kerr microscopy system (with IP and OOP magnet) for observing DW motion.

**Acknowledgements:** The authors gratefully acknowledge the National Research Foundation (NRF), Singapore for the NRF-IIP (NRF2015-IIP003-001) and NRF-CRP (NRF-CRP21-2018-003) grants. The authors also acknowledge the support provided by Agency for Science, Technology and Research, A*STAR RIE2020 AME Grant No. A18A6b0057 for this work. DK acknowledges the financial assistance from NTU research scholarship.